\documentstyle[12pt,epsfig]{article}
\newcommand{\be}{\begin{equation}}
\newcommand{\bea}{\begin{eqnarray}}
\newcommand{\eea}{\end{eqnarray}}
\newcommand{\ba}{\begin{array}}
\newcommand{\ea}{\end{array}}
\newcommand{\ee}{\end{equation}}

\newcommand{\cN}{{\cal N}}

\expandafter\ifx\csname mathbbm\endcsname\relax

\else

\fi

\begin{document}
\begin{titlepage}
\hfill
\vbox{
    \halign{#\hfil         \cr
           CERN-TH/99-266 \cr
           IPM/P-99/045   \cr
           hep-th/9909215   \cr
           } 
      }  
\vspace*{20mm}
\begin{center}
{\Large {\bf  Supergravity and Large $N$ Noncommutative\\ Field Theories}\\} 

\vspace*{15mm}
\vspace*{1mm}
{\ Mohsen Alishahiha$^a$}, 
{\ Yaron Oz$^a$} and 
{\ M.M. Sheikh-Jabbari $^b$\footnote{Address after October 99: ICTP,
Trieste,
Italy}}
\vspace*{1cm} 

{\it $^a$Theory Division, CERN \\
CH-1211, Geneva, 23, Switzerland}\\
\vskip .5cm
{\it $^b$Institute for Studies in Theoretical Physics and Mathematics, \\
 P.O. Box 19395-1795, Tehran, Iran} \\

\vspace*{.5cm}
\end{center}

\begin{abstract}
We consider systems of Dp branes in the presence of a nonzero $B$ field.
We study the corresponding supergravity solutions in the limit
where the branes worldvolume theories decouple from gravity. 
These provide dual descriptions of large $N$ noncommutative field theories.
We analyse
the phase structure of the  theories and the validity of the different description.
We provide evidence 
that in the presence of a nonzero $B$ field the worldvolume theory of D6 branes
decouples from gravity.
We analyse  the systems of M5 branes and NS5 branes in the presence of a 
nonzero  $C$ field and nonzero RR fields, respectively. 
Finally, we study the Wilson loops (surfaces) using the dual 
descriptions.

\end{abstract}
\vskip 4cm

September 1999
\end{titlepage}

\newpage

\section{ Introduction}

The AdS/CFT correspondence (see \cite{OZ} for a review) relates 
field theories without gravity to supergravity (string) theories on certain curved
backgrounds.
The correspondence naturally arises when considering Dp branes in a limit
where the worldvolume field theory decouples from the bulk gravity \cite{Mal}. 
As discussed in \cite{HulDo} and further studied in 
\cite{NCSYM}, when turning on a $B$ field on the D-brane worldvolume the low energy effective 
worldvolume theory is deformed to a noncommutative Super Yang-Mills (NCSYM)
theory.
With  $N$  coinciding Dp branes in the presence of a nonzero $B$ field
the worldvolume theory is deformed to a $U(N)$ NCSYM \cite{SW}. 

Turning on a $B$ field on the D-brane worldvolume can be viewed via the AdS/CFT correspondence
as a perturbation of the worldvolume field theory by a higher dimension operator.
The noncommutative effects are relevant in the UV and are negligible in the IR.
In fact, there is a map from the commutative field theory variables to the noncommutative
ones \cite{SW}.
As in the cases with $B=0$, there exists a limit where the bulk gravity decouples
from the worldvolume noncommutative field theory \cite{CDS,SW}, and a correspondence
between string theory on curved
backgrounds with $B$ field and noncommutative field theories is expected.
The aim of this paper is to study this correspondence using Dp branes, M5 branes 
and NS5 branes. Related works along this directions are \cite{AKI,MALR,Li}.
Other recent studies of noncommutative field theories and string theory are \cite{Others1}.

The paper is organized as follows.
In section 2 we will review the effect of a  $B$ field on the worldvolume
theory of branes.
We will discuss the Dp branes supergravity solutions in the presence of a $B$ field,
the decoupling limit and various aspects of the correspondence with the noncommutative
worldvolume field theories.
We will analyse the phase structure of the Dp branes and plot their phase diagrams.
We will see that the structure can vary depending on the rank of the $B$ field, i.e.
depending on the number of noncommutative coordinates.
We will argue that, unlike the $B=0$ case, in the presence of a nonzero $B$ field
there is a limit where the worldvolume theory
of Dp branes with $p > 5$ decouples from gravity.
In particular, for D6 branes we will see that with two noncommutative coordinates
we have for finite $N$ a UV description in terms of eleven dimensional supergravity
on a curved space. For four or six  noncommutative coordinates
we find  for finite $N$ a UV description in terms of ten dimensional supergravity
on a curved space.

In section 3 we will discuss M5 branes in the presence of a nonzero $C$ field
and NS5 branes in the presence of nonzero RR fields.
In the case of M5 branes wrapping a circle 
we will see the same decoupling limit discussed in \cite{SW}
arising from supergravity. 
However, in the UV the good description of this system is in terms of D4 branes background,
and we do not find a six dimensional field theory description.
Considering  M5 branes with six flat nocompact worldvolume coordinates we   
curiously find another decoupling limit.
At low energies the supergravity background is of the form $AdS_7\times S^4$ with a self-dual
$C$ field which is the dual description of the $(0,2)$ theory. As we increase the energy
the background is deformed and the  $C$ field is no longer self-dual.
In section 4 we will use the dual 
description in order to compute Wilson loops and Wilson surfaces for the
different brane theories.
We will show that, in some cases, in the presence of the nonzero $B$ ($C$) field 
there is way to fix the string (membrane) end point (string) 
by considering a moving coordinates frame in the computation.
Section 5 is devoted to a discussion.


\section{Dp Branes in constant $B$ field}

\subsection{$B$ field background}

Consider string theory in flat space in the background of constant
NS $B$ field and Dp branes. 
In this set up, the end points 
of the open strings attached to the branes, $x_i$, are noncommuting \cite{AAS1}: 
\be
[x_i, x_{i+1}]=il_s^2\; \frac{B_{i,i+1}}{1+B^2_{i,i+1}}|_{on\;the\;brane} \ .
\label{nc}
\ee
We will study this system in the limit $B_{i,i+1} \rightarrow \infty$ and  
$l_s\rightarrow 0$ such that $b_i \equiv l_s^2B_{i,i+1}$ is fixed.
Rescaling the coordinates $x_i \rightarrow \frac{b_i}{l_s^2} x_i$ and keeping the 
new coordinates fixed in the limit we get 
$[x_i, x_{i+1}]=ib_i$.

In the presence of the $B$ field, the massless states excitations of the open strings 
attached to the Dp branes 
give rise to a noncommutative worldvolume field theories, with
$b_i$ being the deformation 
parameters.
The mode 
expansions of the open strings coordinates and momenta are:
\be\ba{cc}
X^i(\sigma,\tau)=x^i+p^i\tau+B^i_{j}p^j \sigma +\; oscil.\;, \\
$\vspace {2mm}$
l_s^2\;P^i(\sigma,\tau)=({\bf 1}-B^2)^i_j(p^j+oscil.),
\ea\ee
where $\sigma,\tau$ parametrize the string world-sheet \cite{AAS1}.
In the above limit the oscillator
modes decouple,
\be
{\bar X}_i(\sigma)\equiv \frac{b_i}{l_s^2}X_i(\sigma)={\bar x}_i+{b_i}{\bar P}_{i+1}
\sigma \ ,
\label{xp}
\ee
where ${\bar P}_i\equiv {l_s^2 \over b_i}P_i$ is rescaled 
in order to preserve the canonical commutator relations.

As we see in (\ref{xp}), there is a finite part added to the string end point, which is
proportional to the momentum. Physically it means that 
the open strings attached to a mixed brane are 
``dipoles'' of the worldvolume $U(N)$ gauge theory 
\cite{{SUSS},{Dip},{YIN}} 
and this, in part, is  a reflection of the non-locality in these theories.
The moment of these dipoles are proportional to $b_iP_{i+1}$.

\subsection{The string (supergravity) description}

In the following we will discuss the dual formulation of noncommutative gauge theories
as string (supergravity) theory on curved backgrounds with a non-zero $B$ field.
Consider now the supergravity description
of Dp branes in the presence of a non-zero $B$ field.
Such solutions were written in
\cite{TSY,AKI,MALR}. It is straightforward to write the most general solutions.
Since we can gauge away the non-zero components of the $B$ field that are normal
to the worldvolume of the branes, the relevant cases are those with 
non-zero components of the $B$ field parallel to the branes.
We denote by $2m$, $ m=1,\cdots, [\frac{p+1}{2}]$, the rank of the $B$ field. The space-time
coordinates are $x_1,...,x_d$ and we denote 
by $x_{p+1}$ the time direction
\footnote{For odd $p$ and when $m=[\frac{p+1}{2}]$ we will consider the 
Euclidean signature. As noted in \cite{MALR}, the decoupling limit   
of the Euclidean and Lorentzian cases are not the same.}.
The supergravity background takes the form \footnote{In the following we will not write
the RR fields.}

\bea
ds^2&=&f_p^{-1/2}\left[\sum_{i~odd
}^{2m-1}h_i(dx_i^2+dx_{i+1}^2)+\cdots+dx_{p+1}^2
\right]+l_s^4f_p^{1/2}(du^2+u^2d\Omega_{8-p}^2),\cr
&&\cr
f_p&=&1+\frac{R^{7-p}}{l_s^4 u^{7-p}},\,\,\,\,\,\,R^{7-p}=c_pg^2_{YM}N
\large(\prod_{i~odd}^{2m-1}\cos\theta_i\large)^{-1},\cr
&&\cr
h_i^{-1}&=&\sin^2 \theta_i f_p^{-1}+\cos^2 \theta_i,\cr
&&\cr
B_{i,i+1}&=&\frac{\sin \theta_i}{\cos \theta_i}f_p^{-1}h_i,\cr
&&\cr
e^{2\phi}&=&g^2f_p^{(3-p)/2}\prod_{i~odd}^{2m-1} h_i \ ,
\label{SG}
\eea
where $c_p = 2^{7-2p}\pi^{\frac{9-3p}{2}}\Gamma(\frac{7-p}{2})$.
The energy coordinate $u$ is related to the radial coordinate $r$ by $u=\frac{r}{l_s^2}$
and $g_{YM}^2=(2\pi)^{p-2}g_s l_s^{p-3}$.

As discussed above, in order to obtain a noncommutative field theory we need to take a 
limit of infinite $B$ field as $l_s \rightarrow 0$.
In this limit we keep fixed the parameters $u,{\bar g}_s, b_i, {\bar x}_{i,i+1}$ defined
by \footnote{For simplicity
we will denote in the rest of the paper the rescaled coordinate
${\bar x}_i$ by $x_i$.} 
\be\ba {ll}
u=\frac{r}{l_s^2}, &{\bar g}_s= g_s l_s^{p-3-2m}, \cr
& \cr
b_i=l_s^2 \tan \theta_i, & {\bar x}_{i,i+1}=\frac{b_i}{l_s^2} x_{i,i+1} \ ,
\label{SCAL}
\ea\ee
where by $x_{i,i+1}$ we mean $x_i,x_{i+1}$.

In the limit (\ref{SCAL}), the supergravity solution (\ref{SG}) 
reads 

\bea
l_s^{-2}ds^2&=&\left(\frac{u}{R}\right)^{{7-p\over2}}\left(\sum_{i~odd}^{2m-1}
h_i(dx_i^2+dx_{i+1}^2)
+\cdots+dx_{p+1}^2\right)+\left(\frac{R}{u}\right )^{{7-p\over2}}(du^2+u^2d\Omega_{8-p}^2),\cr
&&\cr
R^{(7-p)}&=&c_p{\bar g}^2_{YM}N\prod_{i~odd}^{2m-1} b_i,\,\,\,\,\,\,\,\,\,\
a_i^{7-p}=\frac{b_i^2}{R^{(7-p)}}, \cr
&&\cr
B_{i,i+1}&=&\frac{l_s^2}{b_i}\frac{a_i^{7-p}u^{7-p}}{1+a_i^{7-p}u^{7-p}},
\,\,\,\,\,\,\, h_i=\frac{1}{1+a_i^{7-p}u^{7-p}},\cr
&&\cr
e^{2\phi}&=&{\bar g}_s^2 \left ({R\over u}\right )^{(7-p)(3-p)/2}\prod_{i~odd}^{2m-1} \frac{b^2_i}
{1+a_i^{7-p}u^{7-p}} \ ,
\label{NONMET}
\eea
where 
\be
{\bar g}^2_{YM}=(2\pi)^{(p-2)}{\bar g}_s \sim  g_s l_s^{p-3-2m} 
\ee
is the gauge coupling of the noncommutative gauge theory. 

The curvature 
of metric (\ref{NONMET}) in string units 
\be
l_s^2 {\cal R} \sim \frac{1}{g_{eff}} \ ,
\label{R}
\ee
where
$g_{eff}$ is a  dimensionless effective gauge coupling of the noncommutative
field theory given by
\be
g_{eff}^2 \sim {\bar g}_{YM}^2N\prod_{i~odd}^{2m-1}b_iu^{p-3} \ .
\label{geff}
\ee
When $g_{eff} \ll 1$ the perturbative field theory description is valid, while when 
$g_{eff} \gg 1$ the supergravity description is valid.
The $l_s^2 {\cal R}$ expansion corresponds to the strong coupling 
expansion in $\frac{1}{g_{eff}}$ of the noncommutative gauge theory.
We note that the curvature 
of metric (\ref{NONMET}) in string units is proportional, up to a bounded factor, to
the curvature in
string units of the background with $B=0$.

It is convenient to define dimensionless effective non commutativity parameters
\be
a_i^{eff} = a_i u \sim \left(\frac{b_iu^2}{g_{eff}}\right)^{\frac{2}{7-p}},~~~~i=1,3,..,2m-1 \ .
\label{ncp}
\ee
At large distances $L \gg \sqrt{b_i}/g_{eff}$
we have $a_i^{eff}<<1$,
the noncommutative effects are small and
the effective description of the worldvolume theory
is in terms of a commutative field theory.
In this regime the supergravity solutions (\ref{NONMET}) reduce to the low energy
backgrounds considered in 
\cite{ITZ}.
The noncommutativity of the worldvolume theory  
is relevant at distances $L \leq \sqrt{b_i}/g_{eff}$ where $a_i^{eff} \geq  1$. 
The noncommutativity effects can be neglected at energies
\be
u \ll \left({\bar g}_{YM}^2Nb_i^{-1}\prod_{j\neq i}^{2m-1}b_j\right)^{\frac{1}{7-p}},
~~~~~~i=1,3,...,2m-1   \ .
\ee

The effective string coupling $e^{\phi}$ in (\ref{NONMET}) reads
\be
e^{\phi} \sim \frac{g_{eff}^{\frac{7-p}{2}}}{N\prod_{i~odd}^{2m-1} (1 +
(a_i^{eff})^{7-p})^{1/2}} \ .
\label{sc}
\ee
Keeping $g_{eff}$ and $a_i^{eff}$ fixed we see from (\ref{sc}) that $e^{\phi} \sim 1/N$.
Thus the string loop expansion corresponds to the $1/N$ expansion of the noncommutative
gauge theory. 
Note also that
at large $u$ (UV) 
the dilaton in (\ref{NONMET}) 
reads 
\be
e^{\phi} \sim u^{(7-p)(p-2m-3)/4} \ ,
\label{bu}
\ee
which blows up
for $p>2m+3$. 
At small $u$ (IR) the dilaton blows up
for $p<3$ independently of the $B$ field.

We define two scales which will be useful for the discussion in the following sections.
One scale is the energy scale where the effective string coupling is of order one while 
the noncommutative effects are negligible.
It reads
\be
u \sim \left(\frac{N^{\frac{p-3}{7-p}}}
{{\bar g}_{YM}^2\prod_{i~odd}^{2m-1}b_i}\right)^{\frac{1}{p-3}} \ .
\label{udil}
\ee
The second scale is  the energy scale where the effective string coupling is of order one while 
the noncommutative effects are large $a_i^{eff} \gg1$.
It reads
\be
u \sim \left({\bar g}_{YM}^{\frac{14-2p+4m}{3-p+2m}}N\prod_{i~odd}^{2m-1}b_i\right)^
{\frac{1}{7-p}} \ .
\label{udilnc}
\ee

Finally, the supergravity action with the background (\ref{NONMET})
\be
l_s^{-8}\int \sqrt{-g}e^{-2\phi}{\cal R} \sim N^{\frac{p+1}{2}} \ ,
\ee
as for the $B=0$, suggesting that the number of degrees of freedom at large $N$
is the same for the noncommutative and commutative field theories \cite{SUSS}.

\subsection{Phase diagrams}

Summarizing the above discussion,
the effective dimensionless expansion parameters of the Dp branes system in the background of
 non-zero $B$ fields are the number of branes $N$,
the effective gauge coupling $g_{eff}$ and the effective noncommutativity parameters 
$a_i^{eff},i=1,3,...,2m-1$.
For each Dp brane we can plot a phase diagram as a function of these dimensionsless parameters.
Different regions of these phase diagrams will have a good description
in terms of different variables.
Such analysis when $B=0$ was done in \cite{ITZ}.

\vskip .5cm
{\bf D2 branes}
\vskip .5cm

Consider the supergravity solution of $N$ D2-branes in the presence of $B$ field
(\ref{NONMET}).
In this case $m=1$, only the $B_{1 2}$ component is non-zero. Thus,
\bea
ds^2&=&l_s^2\left[{u^{5/2} \over R^{5/2}}\left(-dt^2+{dx_1^2+dx_2^2 \over 
1+a^5u^5}\right)
+\frac{R^{5/2}}{u^{5/2}}(du^2+u^2d\Omega^2_6)\right], \cr
B_{1 2}&=& \frac{l_s^2}{b} \frac{a^5 u^5}{1+a^5u^5},\cr
e^{2\phi}&\sim&\frac{({\bar g}_{YM}^{10}Nb^5)^{1/2}}{u^{5/2}(1+a^5u^5)} \ ,
\label{D2}
\eea
where $B_{1 2}$ is the $B$ field scaled in accord with the coordinates 
rescaling. 
When $a^{eff} \ll 1$ the noncommutativity effects are small and we have 
a good description in terms of a commutative field theory.
This is valid at low energies $u\ll (\frac{{\bar g}_{YM}^2N}{b})^{1/5}$.

\begin{figure}[htb]
\begin{center}
\epsfxsize=6in\leavevmode\epsfbox{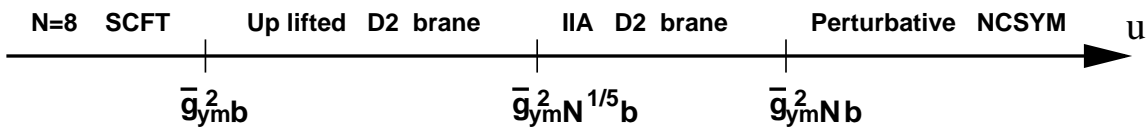}
\end{center}
\caption{The different descriptions of the D2 branes theory with non-zero
$B$ field 
as a function of the energy scale $u$.
We see the flow from $\cN=8$ NCSYM at high energy to $\cN=8$ SCFT 
at low energy. The plot is for the case $\beta \ll1$ and therefor when we up-lift to eleven dimensions
the noncommutativity effects are negligible. When $\beta \gg 1$ the plot is similar, however the transition
to eleven dimensions occurs at $u \sim {\bar g}_{YM}^{14/15}N^{1/5}b^{1/5}$ and then the noncommutative
effects are not negligible.
}
\label{d2fig}
\end{figure}

Consider the flow from high energies to low energies.
The effective dimensionless coupling (\ref{geff}) is now  
$g_{eff}^2 \sim {\bar g}_{YM}^2Nb/u$.
When $g_{eff} \ll 1$ we have a good description in terms of
noncommutative
${\cal N}=8$ perturbative noncommutative super Yang-Mills (NCSYM).
The energy range for this description to be valid
is $u \gg {\bar g}_{YM}^2Nb$.
When $g_{eff} \sim 1$, that is $u\sim {\bar g}_{YM}^2Nb$,
we have a transition to the Type IIA supergravity description.    
The Type IIA supergravity description is valid when both the curvature in string units
(\ref{R}) and the effective string coupling (\ref{sc}) are small.
This implies large $N$ (or large noncommutativity parameter $a^{eff}$).
When the effective string coupling is large the good description is in terms of an
eleven dimensional theory. This description is obtained by uplifting the D2 brane
solution (\ref{D2}) to eleven dimensions.
When uplifting to eleven dimensions we can distinguish two cases.
In the first case the effective string coupling becomes large before the
noncommutative effects can be neglected while in the second case it becomes
large after the noncommutative effects become negligible.
It is convenient to define a dimensionless parameter $\beta$ which is the ratio between
the energy scale at which the effective string coupling is of order one while the
noncommutative effects are negligible
and the energy scale at which 
the dimensionless noncommutative parameter $a^{eff}$ is of order one.   
It reads
$\beta = {\bar g}_{YM}^4 b^3$.
Then the first case corresponds to $\beta \gg 1$ and the second case to
$\beta \ll 1$.
Finally, at energies $u \ll  {\bar g}_{YM}^2 b$ the good description is
in terms of the eleven dimensional M2 branes background.
In figure \ref{d2fig} we plot the transition between the different descriptions
as a function of the energy scale $u$.
We see the flow from $\cN=8$ NCSYM at high energy to $\cN=8$ SCFT 
at low energy.   

\vskip .5cm
{\bf D4 branes}
\vskip .5cm

We will consider now $N$ D4 branes in the presence of a non-zero $B$ field. 
The rank $2m$ of the
$B$ field can be two or four.

\vskip .3cm

{\large $m=1$}
\vskip .3cm

\begin{figure}[htb]
\begin{center}
\epsfxsize=6in\leavevmode\epsfbox{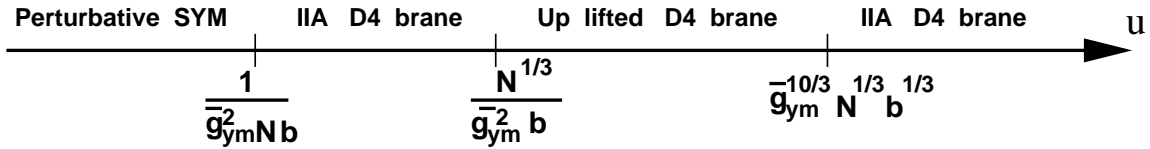}
\end{center}
\caption{The different descriptions of the D4 branes theory with 
$B$ field ($m=1$)
as a function of the energy scale $u$ for $\beta \ll 1$.
}
\label{d4fig}
\end{figure}

\begin{figure}[htb]
\begin{center}
\epsfxsize=4in\leavevmode\epsfbox{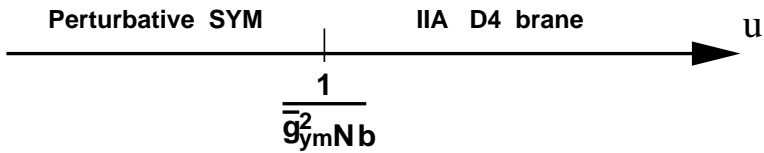}
\end{center}
\caption{The different descriptions of the D4 branes theory with 
$B$ field ($m=1$)
as a function of the energy scale $u$ for $\beta \gg 1$.
}
\label{d41fig}
\end{figure}

When $a^{eff} \ll 1$ the noncommutativity effects are small and we have 
a good description in terms of a commutative field theory.
This is valid at low energies $u\ll (\frac{{\bar g}_{YM}^2N}{b})^{1/3}$.
Consider the flow from low energies to high energies.
The effective dimensionless coupling (\ref{geff}) is now  
$g_{eff}^2 \sim {\bar g}_{YM}^2Nbu$.
When $g_{eff} \ll 1$ we have a good description in terms of
a maximally supersymmetric five dimensional Yang-Mills theory.
The energy range for this description to be valid
is $u \ll \frac{1}{ {\bar g}_{YM}^2Nb}$.
When $g_{eff} \sim 1$, that is $u\sim \frac{1}{ {\bar g}_{YM}^2Nb}$
we have a transition to the Type IIA supergravity description.    
The Type IIA supergravity description is valid when both the curvature in string units
(\ref{R}) and the effective string coupling (\ref{sc}) are small.
This implies large $N$ or large noncommutativity parameter $a^{eff}$.
When the effective string coupling is large the good description is in terms of an
eleven dimensional theory. This description is obtained by uplifting the D4 brane
solution to eleven dimensions.
As in the D2 brane case, when uplifting to eleven dimensions we can distinguish two cases.
In the first case the effective string coupling becomes large before the
noncommutative effects become significant while in the second case it becomes
large after the noncommutative effects become significant.
The the ratio between
the energy scale at which the effective string coupling is of order one
and the energy scale at which 
the dimensionless noncommutative parameter $a^{eff}$ is of order one reads now   
$\beta = \frac{1}{{\bar g}_{YM}^4 b}$.
The first case corresponds to $\beta \ll 1$ and the second case to
$\beta \gg 1$.
When $\beta \ll 1$ we up lift to eleven dimensions at energy 
$u \sim \frac{N^{1/3}}{{\bar g}_{YM}^2 b}$.
As we increase the energy the noncommutative effects become large and the  
effective string coupling decreases. It becomes small again at energies
$u\gg {\bar g}_{YM}^{10/3}N^{1/3}b^{1/3}$ and we have a good description by the Type IIA 
supergravity background.
In figure \ref{d4fig} we plot the transition between the different descriptions
as a function of the energy scale $u$.
Finally, when $\beta \gg 1$ we do not have to up lift to eleven dimensions.
The reason being that the 
effective string coupling is kept small by the large noncommutative effects.
This is described in figure 
\ref{d41fig}.

\vskip .3cm

{\large $m=2$}
\vskip .3cm

The case $m=2$ is similar to to the $m=1$ case and we will briefly discuss
it. 
For simplicity consider the case $b_1=b_3=b$. 
It is again convenient to define the dimensionless parameter $\beta$ which
now reads $\beta = \frac{1}{{\bar g}_{YM}^4 b^3}$.
The phase diagram for the cases $\beta \gg 1$ and $\beta \ll 1$ are similar to the $m=1$ case
above.
The energy scales at which the transitions occur are, of course, modified.

\vskip .3cm
{\bf D5 branes}
\vskip .3cm

Consider now the theory of 
$N$ D5 branes of  Type IIB string theory in the presence of a $B$ field. 
The rank of the B-field can be up to six,
$m=1,2,3$.

\vskip .3cm

{\large $m=1$}
\vskip .3cm

\begin{figure}[htb]
\begin{center}
\epsfxsize=4.5in\leavevmode\epsfbox{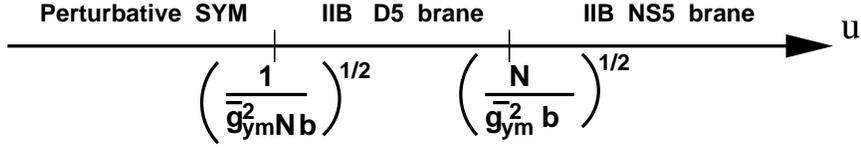}
\end{center}
\caption{The transition between the different descriptions of the D5 brane theory
with $B$ field ($m=1$) as a function of the energy scale $u$ when 
$\beta \ll 1$.}
\label{d5fig}
\end{figure}

\begin{figure}[htb]
\begin{center}
\epsfxsize=4in\leavevmode\epsfbox{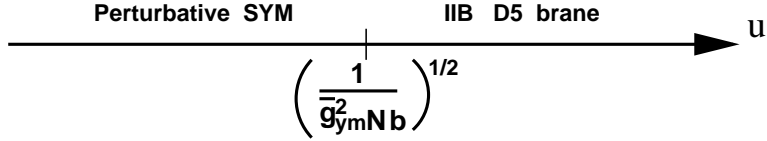}
\end{center}
\caption{The transition between the different descriptions of the D5 brane theory
with $B$ field ($m=1$)
as a function of the energy scale $u$ when 
$\beta \gg 1$.}
\label{d51fig}
\end{figure}

The noncommutativity effects are small and we have 
a good description in terms of a commutative field theory 
at low energies $u\ll (\frac{{\bar g}_{YM}^2N}{b})^{1/2}$.
Consider the flow from low energies to high energies.
The effective dimensionless coupling (\ref{geff}) is now  
$g_{eff}^2 \sim {\bar g}_{YM}^2Nbu^2$.
When $g_{eff} \ll 1$ we have a good description in terms of
a maximally supersymmetric six dimensional Yang-Mills theory.
The energy range for this description to be valid
is $u \ll (\frac{1}{ {\bar g}_{YM}^2Nb})^{1/2}$.
When $g_{eff} \sim 1$, that is $u\sim (\frac{1}{ {\bar g}_{YM}^2Nb})^{1/2}$
we have a transition to the Type IIB supergravity description.    
The Type IIB supergravity description is valid when both the curvature in string units
(\ref{R}) and the effective string coupling (\ref{sc}) are small.
As before, this implies large $N$ or large noncommutativity parameter $a^{eff}$.
When effective string coupling is large the good description is in terms of an
S-dual ten dimensional theory.
We distinguish two cases.
In the first case the effective string coupling becomes large before the
noncommutative effects become significant while in the second case it becomes
large after the noncommutative effects become significant.
The the ratio between
the energy scale at which the effective string coupling is of order one
and the energy scale at which 
the dimensionless noncommutative parameter $a^{eff}$ is of order one reads now   
$\beta = \frac{1}{{\bar g}_{YM}^2}$.
The first case corresponds to $\beta \ll 1$ and the second case to
$\beta \gg 1$.
When $\beta \ll 1$ we use the S-dual description when
$u \sim (\frac{N}{{\bar g}_{YM}^2b})^{1/2}$ .
As we increase the energy the noncommutative effects become large and the  
effective string coupling approaches the value $\frac{1}{\beta}={\bar g}_{YM}^2$. 
In figure \ref{d5fig} we plot the transition between the different descriptions
as a function of the energy scale $u$.
When $\beta \gg 1$ the 
effective string coupling is kept small by the large noncommutative effects and
we do not need the S-dual description.
This is described in figure 
\ref{d51fig}.

\vspace*{.3cm}

{\large $m=2$}

\vskip .3cm

\begin{figure}[htb]
\begin{center}
\epsfxsize=5in\leavevmode\epsfbox{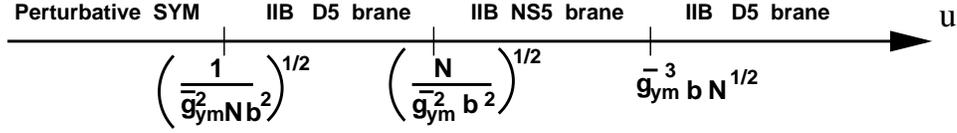}
\end{center}
\caption{The transition between the different descriptions of the D5 brane theory
with $B$ field ($m=2$) as a function of the energy scale $u$ when 
$\beta \ll 1$.}
\label{d52fig}
\end{figure}

\begin{figure}[htb]
\begin{center}
\epsfxsize=4in\leavevmode\epsfbox{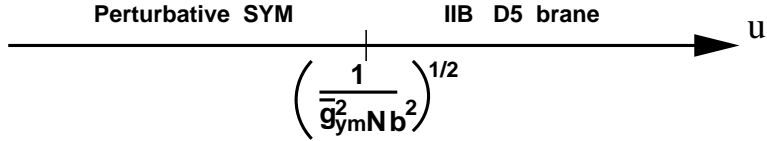}
\end{center}
\caption{The transition between the different descriptions of the D5 brane theory
with $B$ field ($m=2$)
as a function of the energy scale $u$ when 
$\beta \gg 1$.}
\label{d521fig}
\end{figure}

For a simplicity of the discussion we will assume $b_1=b_3=b$. 
The noncommutativity effects are small and we have 
a good description in terms of a commutative field theory 
at low energies $u\ll ({\bar g}_{YM}^2N)^{1/2}$.
Consider the flow from low energies to high energies.
The effective dimensionless coupling (\ref{geff}) is now  
$g_{eff}^2 \sim {\bar g}_{YM}^2Nb^2u^2$.
When $g_{eff} \ll 1$ we have a good description in terms of
a maximally supersymmetric six dimensional Yang-Mills theory.
The energy range for this description to be valid
is $u \ll (\frac{1}{ {\bar g}_{YM}^2Nb^2})^{1/2}$.
When $g_{eff} \sim 1$, that is $u\sim (\frac{1}{ {\bar g}_{YM}^2Nb^2})^{1/2}$
we have a transition to the Type IIB supergravity description.    
When the effective string coupling is large we have to pass to the
S-dual description.
As in the previous analysis, we distinguish two cases.
In the first case the effective string coupling becomes large before the
noncommutative effects become significant while in the second case it becomes
large after the noncommutative effects become significant.
The the ratio between
the energy scale at which the effective string coupling is of order one
and the energy scale at which 
the dimensionless noncommutative parameter $a^{eff}$ is of order one reads now   
$\beta = \frac{1}{{\bar g}_{YM}^2b}$.
The first case corresponds to $\beta \ll 1$ and the second case to
$\beta \gg 1$.
When $\beta \ll 1$ we use the S-dual description when
$u \sim (\frac{N}{{\bar g}_{YM}^2b^2})^{1/2}$ .
As we increase the energy the noncommutative effects become large and the  
effective string coupling decreases.
At energy scales $u \gg {\bar g}_{YM}^3bN^{1/2}$ we can use the Type IIB description
again.
In figure \ref{d52fig} we plot the transition between the different descriptions
as a function of the energy scale $u$.
When $\beta \gg 1$ the 
effective string coupling is kept small by the large noncommutative effects and
we do not need the S-dual description.
This is described in figure 
\ref{d521fig}.

\vspace*{.3cm}

{\large $m=3$}
\vskip .3cm

The case of $m=3$ is similar to to the $m=2$  case and we will briefly discuss
it. 
We consider the 
Euclidean signature and again 
assume $b_1=b_3=b_5=b$. 
It is again convenient to define the dimensionless parameter $\beta$ which
now reads $\beta = \frac{1}{{\bar g}_{YM}^2b}$.
The phase diagram for the cases $\beta \gg 1$ and $\beta \ll 1$ are similar to the $m=2$ case
above.
The energy scales at which the transitions occur are modified.

\vskip .3cm
{\bf D6 branes}
\vskip .3cm

With a vanishing $B$ field the worldvolume theory of $N$ D6 branes of Type IIA
string theory does not decouple from the bulk.
This can be seen, for instance, by the fact that in the decoupling limit
we keep $g_{YM}^2 = g_sl_s^3 = fixed$ as $l_s \rightarrow 0$. This means that
the eleven dimensional Planck length $l_p= g_s^{1/3}l_s$ is kept fixed and that gravity 
does not decouple.

Consider now $N$ D6 branes of Type IIA in the
presence of a $B$ field. 
In this case the rank of the $B$ field can be up to six,
$m=1,2,3$. 
The effective string coupling (\ref{bu}) at large $u$ reads
$e^{\phi} \sim u^{(3-2m)/4}$. When $m=1$ we expect to have an eleven dimensional description 
in the UV.
Note that in the decoupling limit we keep $g_sl_s^{3-2m}=fixed$ as
$l_s\rightarrow 0$.
Therefor for $m=1$ the 
the eleven dimensional Planck length $l_p\rightarrow 0$ and we expect gravity to decouple.
For $m=2,3$ the effective string coupling is small at all energy scales
and there is no need for an eleven dimensional description at high energy.
The ten dimensional Planck scale $l_p^{(10)}= g_s^{1/4}l_s\rightarrow 0$ 
and we expect gravity to decouple.
In the following we will analyse the phase diagram of the D6 branes system.

The background in the limit (\ref{SCAL}) takes the form 
\bea
l_s^{-2}ds^2&=&\frac{u^{1/2}}{R^{1/2}}\left(\sum_{i~odd}^{2m-1}h_i
(dx_i^2+dx_{i+1}^2)
+\cdots+dx_{7}^2\right)+\frac{R^{1/2}}{u^{1/2}}(du^2+u^2d\Omega_{2}^2),\cr
&&\cr
R&=&c_p{\bar g}^2_{YM}N\prod_{i~odd}^{2m-1} b_i,\,\,\,\,\,\,\,\,\,\,\,\,\,\,
a_i=\frac{b_i^2}{R}, \cr
&&\cr
B_{i,i+1}&=&\frac{l_s^2}{b_i}\frac{a_iu}{1+a_iu},
\,\,\,\,\,\,\,\,\,\, h_i=\frac{1}{1+a_iu},\cr
&&\cr
e^{2\phi}&\sim&\left(\frac{{\bar g}_{YM}^2\prod_i b_i}{N^3}\right)^{1/2} 
u^{3/2}\prod_{i~odd}^{2m-1} \frac{1}{1+a_iu} \ .
\eea
\vskip .5cm

{\large $m=1$}
\vskip .3cm

The noncommutativity effects are small and we have 
a good description in terms of a commutative field theory 
at low energies $u\ll \frac{{\bar g}_{YM}^2N}{b}$.
Consider the flow from low energies to high energies.
The effective dimensionless coupling (\ref{geff}) is now  
$g_{eff}^2 \sim {\bar g}_{YM}^2Nbu^3$.
When $g_{eff} \ll 1$ we have a good description in terms of
a perturbative maximally supersymmetric seven dimensional Yang-Mills theory.
The energy range for this description to be valid
is $u \ll (\frac{1}{ {\bar g}_{YM}^2Nb})^{1/3}$.
When $g_{eff} \sim 1$, that is $u\sim (\frac{1}{ {\bar g}_{YM}^2Nb})^{1/3}$
we have a transition to the Type IIA supergravity description.    
When effective string coupling is large the good description is in terms of an
eleven dimensional theory.
As before, we distinguish two cases.
In the first case the effective string coupling becomes large before the
noncommutative effects become significant while in the second case it becomes
large after the noncommutative effects become significant.
The the ratio between
the energy scale at which the effective string coupling is of order one
and the energy scale at which 
the dimensionless noncommutative parameter $a^{eff}$ is of order one reads now   
$\beta = \frac{b}{{\bar g}_{YM}^4}$.
The first case corresponds to $\beta \ll 1$ and the second case to
$\beta \gg 1$.
When $\beta \ll 1$ we use the eleven dimensional supergravity description when
$u \sim \frac{N}{({\bar g}_{YM}^2b)^{1/3}}$ .
The eleven dimensional curvature is small for every  $N$ when 
$u >  \frac{N}{({\bar g}_{YM}^2b)^{1/3}}$ 
\be
l_p^2 {\cal R}_{11} \sim e^{2\phi/3} \frac{1}{g_{eff}} < 1/N^2 \ ,
\ee
and vanishes for $u \gg  \frac{N}{({\bar g}_{YM}^2b)^{1/3}}$.Thus,
similar to the case without a $B$ field \cite{ITZ}, the eleven dimensional supergravity solution
can be trusted in the UV for any $N$.  
Unlike the $B=0$ case, the metric at large $u$ is not the flat eleven dimensional one.
As we discussed above, since the eleven dimensional Planck length goes to zero in
the decoupling limit we expect gravity
to decouple from the branes worldvolume theory.
Thus, it is plausible that a seven dimensional worldvolume theory 
without gravity does exist.

\begin{figure}[htb]
\begin{center}
\epsfxsize=5in\leavevmode\epsfbox{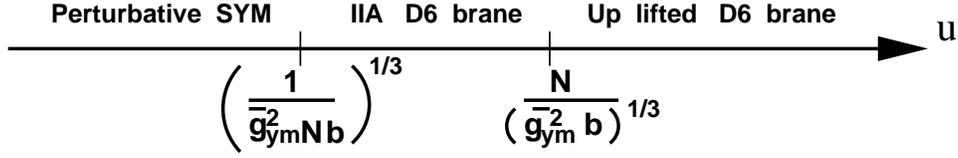}
\end{center}
\caption{The transition between the different descriptions of the D6 brane theory
with $B$ field ($m=1$)
as a function of the energy scale $u$ when 
$\beta \ll 1$.
When $\beta \gg 1$ the plot is similar. However, the transition to the eleven dimensional
description is at $u \sim \frac{Nb}{{\bar g}_{YM}^6}$.}
\label{d6fig}
\end{figure}

When $\beta \gg 1$ the phase diagram is similar, however the transition
to eleven dimensions occurs at $u \sim \frac{Nb}{{\bar g}_{YM}^6}$ 
and then the noncommutative
effects are not negligible.
Similarly, the eleven dimensional supergravity solution
can be trusted in the UV for any $N$.  
In figure \ref{d6fig} we plot the transition between the different descriptions
as a function of the energy scale $u$.

\vspace*{.3cm}

{\large $m=2$}
\vskip .3cm

\begin{figure}[htb]
\begin{center}
\epsfxsize=6in\leavevmode\epsfbox{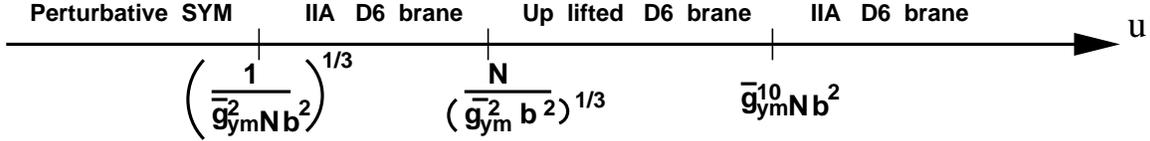}
\end{center}
\caption{The transition between the different descriptions of the D6 brane theory
with $B$ field ($m=2$) as a function of the energy scale $u$ when 
$\beta \ll 1$.}
\label{d61fig}
\end{figure}

For a simplicity of the discussion we will assume $b_1=b_3=b$. 
The noncommutativity effects are small and we have 
a good description in terms of a commutative field theory 
at low energies $u\ll {\bar g}_{YM}^2N$.
Consider the flow from low energies to high energies.
The effective dimensionless coupling (\ref{geff}) is now  
$g_{eff}^2 \sim {\bar g}_{YM}^2Nb^2u^3$.
When $g_{eff} \ll 1$ we have a good description in terms of
perturbative seven dimensional Yang-Mills theory.
The energy range for this description to be valid
is $u \ll (\frac{1}{ {\bar g}_{YM}^2Nb^2})^{1/3}$.
When $g_{eff} \sim 1$, that is $u\sim (\frac{1}{ {\bar g}_{YM}^2Nb^2})^{1/3}$
we have a transition to the Type IIA supergravity description.    
As in the previous analysis, we distinguish two cases.
In the first case the effective string coupling becomes large before the
noncommutative effects become significant while in the second case it becomes
large after the noncommutative effects become significant.
The the ratio between
the energy scale at which the effective string coupling is of order one
and the energy scale at which 
the dimensionless noncommutative parameter $a^{eff}$ is of order one reads now   
$\beta = \frac{1}{{\bar g}_{YM}^4b}$.
The first case corresponds to $\beta \ll 1$ and the second case to
$\beta \gg 1$.
When $\beta \ll 1$ we use the eleven dimensional description when
$u \sim \frac{N}{({\bar g}_{YM}^2b^2)^{1/3}}$ .
As we increase the energy the noncommutative effects become large and the  
effective string coupling decreases.
At energy scales $u \gg {\bar g}_{YM}^{10}Nb^2$ we can use the Type IIA description
again.
The ten dimensional curvature is small for every  $N$ when 
$u > {\bar g}_{YM}^{10}Nb^2$ 
\be
l_s^2 {\cal R}_{10} < 1/N^2 \ ,
\ee
and vanishes for $u \gg {\bar g}_{YM}^{10}Nb^2$.Thus,
the ten  dimensional supergravity solution
can be trusted in the UV for any $N$.  
Note, however, the metric at large $u$ is not flat. 
In figure \ref{d61fig} we plot the transition between the different descriptions
as a function of the energy scale $u$.

\begin{figure}[htb]
\begin{center}
\epsfxsize=4in\leavevmode\epsfbox{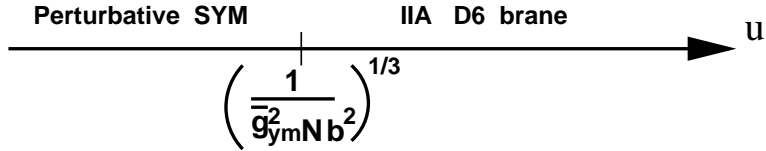}
\end{center}
\caption{The transition between the different descriptions of the D6 brane theory
with $B$ field ($m=2$)
as a function of the energy scale $u$ when 
$\beta \gg 1$.}
\label{d62fig}
\end{figure}

When $\beta \gg 1$ the 
effective string coupling is kept small by the large noncommutative effects and
we do not need the eleven dimensional description.
The ten dimensional curvature is small for every  $N$ when 
$u > \frac{N}{({\bar g}_{YM}^2b^2)^{1/3}}$
and the ten  dimensional supergravity solution
can be trusted in the UV for any $N$.  
This is described in figure 
\ref{d62fig}.
The interaction Lagrangian between the brane modes and the bulk modes
is proportional to positive powers of $\kappa_{10} = g_sl_s^4$ which goes to zero in
the decoupling limit. Thus we expect all the interaction terms to vanish
in this limit and gravity
to decouple from the branes worldvolume theory. 
Thus, it is plausible to expect that a seven dimensional worldvolume theory 
without gravity does exist. 
It was noted in \cite{MALR} that such a theory will have a negative specific heat.

\vspace*{.3cm}

{\large $m=3$}
\vskip .3cm

The case $m=3$ is similar to to the $m=2$  case and we will briefly discuss
it. 
We 
assume $b_1=b_3=b_5=b$. 
It is again convenient to define the dimensionless parameter $\beta$ which
now reads $\beta = \frac{1}{{\bar g}_{YM}^4b^3}$.
The phase diagram for the cases $\beta \gg 1$ and $\beta \ll 1$ are similar to the $m=2$ case
above.
The energy scales at which the transitions occur are modified. 
As for the $m=2$ case, the scalar curvature vanishes at large $u$,
however, the metric at large $u$ is not flat. 
The decoupling from the bulk argument is as in the $m=2$ case.

\vskip .3cm
{\bf Dp branes ($p >6$)}
\vskip .3cm

Consider now the decoupling limit for Dp branes with $p>6$.
In a ten dimensional description 
the interaction Lagrangian between the brane modes and gravity
is proportional to positive powers of $\kappa_{10} = g_sl_s^4$.
In an eleven dimensional description the interaction
is proportional to positive powers of the eleven dimensional Planck length
$l_p$.
Consider first the D7 branes. In the decoupling limit we hold $g_sl_s^{4-2m}$ fixed as 
$l_s \rightarrow 0$.
Therefor, when $m>0$ we see that  $\kappa_{10} \rightarrow 0$ in this limit, 
which indicates that the worldvolume theory decouples from gravity.   
In the D8 branes case we hold $g_sl_s^{5-2m}$ fixed as 
$l_s \rightarrow 0$.
Again, when $m>0$ we see that  $\kappa_{10} \rightarrow 0$ in this limit.
When $m=1$ the effective string coupling 
 is small in the UV and the ten dimensional description is  sufficient.
When $m>1$ the effective string coupling is large in the UV
and we will need an  eleven dimensional description.
Note that $l_p \rightarrow 0$ when $m>1$ which indicates that again gravity decouples
from the brane worldvolume theory.
For D9 branes we  hold $g_sl_s^{4-2m}$ fixed as 
$l_s \rightarrow 0$ which ensures again that  $\kappa_{10} \rightarrow 0$
and indicates the decoupling of gravity.

\subsection{Non Extremal Dp Branes}

Consider the non-extremal Dp branes solution with non zero $B$ field. 
The metric in the decoupling
limit reads:
\bea
l_s^{-2}ds^2=
&&\left(\frac{u}{R}\right)^{{7-p\over2}}\left(\sum_{i~odd}^{2m-1}h_i(dx_i^2+dx_{i+1}^2)
+\cdots+(1-(\frac{u_T}{u})^{7-p})dx_{p+1}^2\right)+\nonumber\\
&&\left(\frac{R}{u}\right)^{{7-p\over 2}}\left(\frac{du^2}{1-(\frac{u_T}{u})^{7-p}}
+u^2d\Omega_{8-p}^2
\right) \ ,
\eea
where $u_T$ is related to the energy density of the brane above density extremality $\varepsilon$
by
\be
u_T^{7-p} \sim {\bar g}_{YM}^4\prod_{i~odd}^{2m-1}b_i^2 \varepsilon \ .
\ee
This should correspond to decoupled theories at finite temperature with
$\varepsilon$ being the energy density of the field theory.
As discussed in \cite{MALR} the thermodynamic quantities are as in 
the case without a $B$ field.
More precisely, they are the same as in the $B=0$ case with
$g_{YM}^2 \rightarrow {\bar g}^2_{YM}\prod_{i~odd}^{2m-1} b_i$.
Later we will analyse the Wilson loops of this system.

\section{Fivebranes}

In this section we will discuss possible noncommutative deformations
of the M5 branes and NS5 branes worldvolume theories. 
\subsection{M5 branes}

Consider $N$ coinciding M5 branes in the presence of a nonzero $C$ field with
$m=1,2$.
The supergravity solution reads \footnote{In the following we will not write the 
component of the $C$ field associated with the M5 branes charge.}

\bea
ds^2_{11}&=&f^{-1/3}\left[ (\prod_{i~odd}^{2m-1} h_i)^{-1/3}\left(\sum_{i~odd}^{2m-1}
 h_i(dx_i^2+dx_{i+1}^2)
+dx_6^2 \right) + (\prod_{i~odd}^{2m-1} h_i)^{2/3}dx_5^2 \right]\cr
&+& f^{2/3}(\prod_{i~odd}^{2m-1} h_i)^{-1/3} (dr^2+r^2d\Omega^2_4) \ ,\cr
&&\cr
f&=&1+\frac{\pi N l_p^3}{\prod_i \cos \theta_i r^3} \ ,\cr
&&\cr
h_i^{-1}&=&\sin^2\theta_i f^{-1}+\cos^2\theta_i \ ,\cr
&&\cr
C_{5,i,i+1}&=&\tan \theta_i\; f^{-1}\ ,\;h_i \;\;\;\;\;\;\;C_{i,i+1,6}=
\sin \theta_{4-i}\cos \theta_{i}\;f^{-1}\;h_i \ .
\label{M5M}
\eea

Let us discuss first the case when the worldvolume
coordinate $x_5$ is compactified on a circle 
of radius
$R_0$.
In the decoupling limit we send $l_p\rightarrow 0$ and 
keep following quantities fixed:
\be\ba{ll}
u=\frac{r}{l_p^{3}}\;R_0\;\;\;\;& {\bar R}_0=\frac{R_0}{l_p^{3m/(m+1)}}\cr
& \cr
 b_i=\frac{l_p^{3}}{R_0} \tan \theta_i\;\;\;& {\bar x}_{i,i+1}=\frac{R_0 b_i}
{l_p^{3}} x_{i,i+1}\cr
&\cr
{\bar x}_6=x_6\;\;\; & {\bar x}_5=\frac{\prod_i b_i R_0^m}{l_p^{3m}} x_5 \ .
\label{scalem5}
\ea\ee
This decoupling limit is consistent with the D4 branes decoupling limit 
where we use the relation $l_s^2R_0 = l_p^3$.
The same scaling of the coordinates $x$ was derived in \cite{SW}
for the case of $m=2$ and $b_i=b$.

In the limit (\ref{scalem5}), the supergravity solution reads:

\bea
l_p^{-2}ds^2_{11}&=&(\prod_i h_i)^{1/3}\frac{u}{(\pi N)^{1/3}\prod_i b_i
{\bar R}_0^{m+1}}
\left(\sum_i h^{-1}_i(dx_i^2+dx_{i+1}^2)
+dx_6^2 + \prod_i h^{-1}_idx_5^2 \right)\cr
&+&(\prod_i h_i)^{1/3} \frac{(\pi N)^{2/3}}{u^2}(du^2+u^2d\Omega^2_4), \cr
&&\cr
h_i&=&1+a_i^3u^3, \;\;\;\;\;\;\;\;\;
a_i^3=\frac{b_i^2}{\pi N {\bar R}_0^{m+1}\prod_j b_j} \ .
\label{m5sol}
\eea
The $C$ field (up to numerical factors) takes the form

\be\ba {lll}
m=1:&
C_{346} \sim \frac{l_p^3}{b^2{\bar R}_0^2}a^3u^3,&
C_{125} \sim \frac{l^3_p}{b^2{\bar R}_0^2}\;\frac{a^3u^3}{1+a^3u^3},\cr
m=2:&
C_{i,i+1,6} \sim \frac{l_p^3}{ b_i^3{\bar R}_0^3}\;\frac{a_i^3u^3}{1+a_i^3u^3},&
C_{i,i+1,5} \sim \frac{l_p^{3}}{ b_i\prod_j b_j {\bar R}_0^3}\;\frac{a_i^3u^3}
{1+a_i^3u^3} \ .
\ea\ee
This background is
the ten dimensional D4 branes solution up lifted to eleven dimensions. 
At low energies compared to $1/R_0$ the description of the system is in terms
of the D4 brane theory, as discussed in the previous section.
We might have expected that at high energies (large $u$) we will have a good description in
terms of a noncommutative $(0,2)$ theory in six dimensions.
The curvature reads
\be
l_p^2{\cal R}_{11}\sim \frac{1}{N^{2/3}\prod_{i~odd}^{2m-1} (1+a_i^3u^3)^{1/3}} \ ,
\ee
and we can trust the supergravity solution.
However, the size of the compact direction, $x_5$, is controlled
$\prod_{i~odd}^{2m-1} h_i^{-2/3}\frac{u}{N^{1/3}}$ which decreases in the UV.
Therefor, at large $u$ we are back in the ten dimensional D4 branes background, as discussed
in the previous section and we do not find a six dimensional field theory description
of the UV.

Let us discuss now the M5 branes in the background of a nonzero $C$ field
without wrapping a circle.
Consider the supergravity solution (\ref{M5M}) and let us keep the following quantities
fixed as $l_p \rightarrow 0$
\be
u^{n-1}= \frac{r}{l_p^n},\;\;\;\;\; b_i^{q/2} = l_p^q \tan \theta_i \ .
\ee
For the moment we will consider $n>1,q$ as arbitrary positive integers.
We get
\bea
f&=&1+\frac{\pi N \prod_i b_i^{q/2}}{l_p^{mq+3n-3}u^{3n-3}} \ ,\cr
&&\cr
h_i&=&\frac{b_i^q}{l_p^{2q}}\;\frac{1}{1+\frac{b_i^q}{l_p^{2q}}(
1+\frac{\pi N \prod_i b_i^{q/2}}{l_p^{mq+3n-3}u^{3n-3}})^{-1}} \ .
\eea
The condition
for a finite metric solution and a  
constant nonzero $C$ field at infinity require $(m-2)q \leq 3(1-n) < mq$.
Keeping finite the tension of the strings that arise from M2 branes  stretched between
the M5 branes requires $n=3$, namely $u^2= \frac{r}{l_p^3}=fixed$.
This implies $m=1,q=6$.
The background reads 
\bea
l_p^{-2}ds_{11}^2&=&\frac{u^2}{(\pi
N)^{1/3}}h^{1/3}(h^{-1}dx^2_{1,2,5}+dx^2_{3,4,6})+\frac{(\pi
N)^{2/3}}{u^2}h^{1/3}(4du^2+u^2d\Omega_4^2) \ ,\cr 
&&\cr
h&=&1+a^6u^6\ ,\cr 
&&\cr
C_{346}&=&\frac{l_p^{3}}{b^{3/2}}a^6u^6,\;\;\;\;\;
C_{125}=\frac{l^{3}_p}{b^{3/2}}\;\frac{a^6u^6}{1+a^6u^6} \ ,
\label{m5new}
\eea
where $a^6=\frac{b^3}{\pi N}$ and we rescaled the coordinates 
$x_{3,4,6}\rightarrow \frac{l_p^3}{b^{3/2}}x_{3,4,6}$ and
$x_{1,2,3}\rightarrow \frac{b^{3/2}}{l_p^3}x_{1,2,3}$.
Note that the decoupling limit leading to (\ref{m5new}) differs from (\ref{scalem5}).

At very low energies (small $u$)  the metric (\ref{m5new}) describes the eleven dimensional
$AdS_7 \times S^4$ background with a self-dual $C$ field,
providing a dual description of the $(0,2)$ SCFT.
As we increase $u$ the $AdS_7 \times S^4$ is deformed and the $C$ field is no longer self-dual.
The curvature reads
\be
l_p^2{\cal R}_{11}\sim \frac{1}{N^{2/3}(1+a^6u^6)^{1/3}} \ ,
\ee
and we can trust the supergravity solution in the UV as well.

\subsection{NS5 branes}

\vskip .3cm
{\bf Type IIB}
\vskip .3cm

The Type IIB NS5 branes solution in the presence of nonzero RR
fields can be obtained from D5 branes by S-duality transformation. Under
S-duality we have:
\be\ba {ll}
l_s^2\rightarrow {l'}_s^2 \equiv g_sl_s^2&g_s \rightarrow g'_s\equiv\frac{1}{g_s} \ ,\cr
e^{\phi}\rightarrow e^{\phi'}\equiv e^{-\phi}&ds^2\rightarrow ds'^2\equiv g_se^{-\phi}ds^2 \ .
\label{st}
\ea\ee
Using (\ref{st}) we get the Type IIB NS5 branes background,
\bea
{ds'}^2&=&\prod_{i~odd}^{2m-1}h_i^{-1/2}\left[
\sum_{i~odd}^{2m-1}h_i(dx_i^2+dx_{i+1}^2)+\cdots+dx_{6}^2
+f(dr^2+r^2d\Omega_{3}^2)\right],\cr
&&\cr
f&=&1+\frac{c_5 N {l'}_s^2}{\prod_{i~odd}^{2m-1} \cos \theta_i\; r^2}, \cr
&&\cr
h_i^{-1}&=&\sin^2 \theta_i f^{-1}+\cos^2 \theta_i,\cr
&&\cr
e^{2\phi'}&=&{g'}_s^2f\prod_{i~odd}^{2m-1} h_i^{-1} \ ,
\label{NSS}
\eea
and the NS field $B_{ij}$ is mapped to a the RR field $A_{ij}$.
The decoupling limit is derived by applying (\ref{st}) on
the decoupling limit of the D5 branes.
It is defined by taking the limit $g'_s{l'}_s^2 \rightarrow 0$ 
and keeping fixed
\be\ba {ll}
u=\frac{r}{g'_s{l'}_s^2}&{\bar g'}_s={g'}_s^{-m}{l'}_s^{2-2m} \,\cr
b_i={g'}_s{l'}_s^{2}\tan \theta_i &
{\bar x}_{i,i+1}=\frac{b_i}{g'_s {l'}_s^2}x_{i,i+1} \ .
\label{iib}
\ea\ee
Keeping $u$ fixed means keeping fixed the mass of a D-string stretched between two NS5 branes. 

In the limit (\ref{iib}) the background (\ref{NSS}) reads
\bea
{ds'}^2&=&\frac{{l'}_s^2}{{\bar g'}_s\prod_i b_i}\prod_{i~odd}^{2m-1}h_i^{1/2}\left[
\sum_{i~odd}^{2m-1}h^{-1}_i(dx_i^2+dx_{i+1}^2)+\cdots+dx_{6}^2
+\frac{c_5N{\bar g'}_s}{u^2}\;\prod_{i~odd}^{2m-1} b_i(du^2+u^2d\Omega_{3}^2)\right],\cr
&&\cr
h_i&=&1+a_i^2u^2,\;\;\;\;\;\;\;\;\;\; a_i^2=\frac{b_i^2}{c_5N\prod_{j~odd}^{2m-1}
 b_j {\bar
g'}_s},\;\;\;\;\;\;\;\;\;\;
e^{2\phi'}=\frac{c_5 N}{\prod_{i~odd}^{2m-1} b_i {\bar g'}_s u^2}\prod_{i~odd}^{2m-1} h_i \ .
\label{NSBD}
\eea
The Yang-Mills coupling of the worldvolume theory is $g_{YM}^2 \sim {g'}_s^{-m}{l'}_s^{2-2m}$.
The curvature of the metric reads
\be
{l'}_s^2{\cal R}\sim \frac{1}{N}\;\frac{1}{\prod_i(1+a_i^2u^2)^{1/2}} \ .
\label{RNS}
\ee

When  $a_i^{eff} \equiv a_iu \ll 1$ the supergravity approximation
can be trusted for large N, while when
$a_i^{eff}\gg 1$ the supergravity approximation can be trusted for finite N. 
When 
$m=1$ we see that at large $u$ the effective string coupling 
 is small and we can use the NS5 brane description
in the UV. When $m=2,3$ the effective string coupling is large in the UV 
and we have to use the S-dual description 
of D5 branes.
This is precisely what we saw in the phase structure of the D5 branes system in the
previous section.

\vskip .3cm
{\bf Type IIA}
\vskip .3cm

The background of Type IIA NS5 branes wrapping a circle can be obtained by a 
T-duality transformation \cite{GPR} of the Type IIB NS5 branes.
We compactify the coordinate $x_5$ on a circle and perform
T-duality in $x_5$ on the background  (\ref{NSBD}). 
The decoupled  Type IIA NS5 branes solution reads 
\bea
{ds'}^2&=&\frac{{l'}_s^2}{{\bar g'}_s\prod_{i~odd}^{2m-1} b_i}\prod_{i~odd}^{2m-1}h_i^{1/2}[
\sum_{i~odd}^{2m-1}h^{-1}_i(dx_i^2+dx_{i+1}^2)+\cdots+\prod_{i~odd}^{2m-1}h_i^{-1}dx_5^2
+dx_{6}^2\cr
&&+\frac{c_5N{\bar g'}_s}{u^2}\;\prod_{i~odd}^{2m-1} b_i(du^2+u^2d\Omega_{3}^2)],\cr
&&\cr
h_i&=&1+a_i^2u^2,\;\;\;\;\;\;\;\;\;\; a_i^2=\frac{b_i^2}{c_5N\prod_{j~odd}^{2m-1} b_j {\bar
g'}_s},\;\;\;\;\;\;\;\;\;\; 
e^{2\phi'}=\frac{c_5 N}{\prod_{i~odd}^{2m-1} b_i {\bar g'}_s u^2}\prod_{i~odd}^{2m-1} 
h_i^{1/2} \ ,
\label{NSAD}
\eea
where we have rescaled 
$x_5 \rightarrow \frac{\prod_i b_i}{(g'_s{l'}_s^2)^m}x_5$ and we have taken into account the fact
that under T-duality $\phi \rightarrow \phi-\frac{1}{2}\ln(g_{55})$. 
Note that unlike the Type IIB NS5 branes background where $m=1,2,3$ 
here $m=1,2$. 
The 3-form field $A$ (up to numerical factors) takes the form
\be\ba {lll}
m=1:&
A_{346} \sim \frac{(g'_s{l'}_s^2)^2}{b^2}a^2u^2,&
A_{125} \sim \frac{(g'_s{l'}^2_s)^2}{b^2}\;\frac{a^2u^2}{1+a^2u^2},\cr
m=2:&
A_{i,i+1,6} \sim \frac{(g'_s{l'}^2_s)^3}{ b_i^3}\;\frac{a_i^2u^2}{1+a_i^2u^2},&
A_{i,i+1,5} \sim \frac{(g'_s{l'}_s^2)^{3}}{ b_i\prod_j b_j }\;\frac{a_i^2u^2}
{1+a_i^2u^2} \ .
\ea\ee
The curvature of the metric is the same as for the Type IIB NS5 branes (\ref{RNS}).
In the IR the effective string coupling is large and we have to lift
the solution to eleven dimensions. The background becomes that of wrapped M5
branes. As we increase the energy we can trust the NS5 branes
background which provides a deformation of the wrapped M5 branes background.

\section{Wilson loops}

In this section will use the dual string  
description in order to compute Wilson loops (surfaces) for the
different brane theories.
\subsection{Dp branes}

According to the AdS/CFT correspondence, the expectation value of the
Wilson loop operator of the gauge theory can be computed in the dual string description
by evaluating the partition function of a string whose worldsheet is bounded by
the loop \cite{Malda,rey}.
In the supergravity approximation the dominant contribution comes from the minimal
two dimensional surface bounded by the loop. The expectation value of the
Wilson loop operator is  
\be
\langle W(C)\rangle \sim e^{-S} \ ,
\ee
where $S$ is the string action evaluated on the minimal surface. 
We will use the same prescription in the case of a 
nonzero $B$ field.
The string action reads now
\be
S =\frac{1}{2 \pi l_s^2}\int d\tau d\sigma \sqrt{det g_{ij}}+\frac{1}{2 \pi l_s^2}
\int B_{ij}\partial_{\tau}X^i\partial_{\sigma}X^j ,
\label{WIL}
\ee
where $g_{ij}=\partial_{i}X^{\mu}\partial_{j}X^{\nu} G_{\mu\nu}$ is the induced
metric.

Consider a static $Q{\bar Q}$ configuration. In general the 
quark and antiquark move with velocity ${\bar p}= {b\over l_s^2}p$. 
When $B=0$, in the $l_s\rightarrow 0$ limit,
 the 
velocity appears via  a multiplicative factor in $Q{\bar Q}$ potential,
as expected by the Lorentz symmetry.
When $B\neq 0$ the situation is different. There is no Lorentz symmetry 
and the $B$ field term contributes.
When the strings are not moving the
end points of strings cannot be fixed at a finite distance $L$ from
each other at large $u$ \cite{MALR} since they grow with $u$.
The endpoints of the strings can be fixed at large $u$ as follows.
As was noted in \cite{SUSS}, the interaction of charges of opposite
sign in a magnetic field is nonlocal in the sense that the interaction point in
terms of the 
center of mass coordinate is shifted by a momentum dependent term.
This suggests that  we should use a   moving coordinates frame in the computation
\footnote{We note that in the case $m=\frac{p+1}{2}$ we will not be able to fix 
the end points of the strings at infinity. In this case, the time coordinate $x_{p+1}$
is noncommutative
coordinate. For a static configuration where the potential is time
independent  we cannot find an appropriate shift of the time
coordinate.}.   
Indeed, as seen from (\ref{xp}),
the end points of the open strings attached to the boundary
are quark and anti-quark moving with the same velocity (\ref{xp}).

In the following we will consider Dp branes with $b_i =b, i=1,3,...,2m-1$.
However we will write the final result for arbitrary $b_i$.
We distinguish two cases. In the first case the rank of the $B$ field is not
maximal, thus some of the coordinates
are commutative and the  loop is parametrized by these.
In this case the computation proceeds exactly as in the $B=0$ case.
In the second case the loop is parametrized by the noncommutative coordinates.
We will discuss this case.
We parametrize the string configuration by
$t=\tau, u=\sigma, x_{1}={\bar p}\tau, x_2=x(u)$. 
Equation (\ref{WIL}) reads now
\be
S=\frac{1}{2\pi}\int d\tau du 
\sqrt{(1-h{\bar p}^2)(1+({u\over R})^{7-p}h(\partial_u x)^2)}
+\frac{1}{2\pi}\int d\tau du {{\bar p}\over b}(au)^{7-p}h\partial_u x  \  ,
\label{act}
\ee
where $R$ and $h$ are defined in (\ref{NONMET}).
It is minimized when 
\be
\frac{({u\over R})^{7-p}h(1-h{\bar p}^2)\partial_u x}{{\cal L}_0}
+(au)^{7-p}h{{\bar p}\over b}=const  \ ,
\label{LAGS}
\ee
where ${\cal L}_0$ is the integrand of the first term in (\ref{act}).
 
At large $u$ we have
\be
\frac{({1\over aR})^{7-p}\partial_u x}{\sqrt{1+\frac{(\partial_u x)^2}{(aR)
^{7-p}}}}
+{{\bar p}\over b}=const \ .
\label{const}
\ee
Therefor if we choose the constant in (\ref{const})
to be ${{\bar p}\over b}$ we can 
fix the position of the string at large $u$. With this choice 
equation (\ref{LAGS}) can be solved written as  
\be
\partial_u x={{\bar p}\over b}({u\over R})^{-{7-p\over2}}
\left(({u\over R})^{7-p}-({{\bar p}\over b})^2\right)^{-1/2} \ . 
\ee
Hence 
\be
x(u)=\int_{u_0}^u {{\bar p}\over b}({u\over R})^{-(7-p)}
\left(({u\over R})^{7-p}-({u_0\over R})^{7-p}\right)^{-1/2} \ ,
\ee
where $\partial_u x |_{u_0} \rightarrow \infty$ \footnote{In order for $u_0$ to be
$N$ independent we should take the momentum ${\bar p}$ to be $N$ dependent.}
\be
(au_0)^{7-p} ={\bar p}^2  \ .
\ee

The $Q{\bar Q}$ separation is defined by 
\bea
L=x(u\rightarrow\infty)&=&\int_{u_0}^{\infty}({R\over u_0})^{{7-p\over 2}}
\left(1-({u_0\over u})^{7-p}\right)^{-1/2}({u_0\over u})^{7-p}\cr
&&\cr 
&=&{R^{{7-p\over 2}}\over 7-p} u_0^{{p-5\over 2}} B({1\over2},{6-p\over 7-p}) \ .
\label{LL}
\eea
Using (\ref{act}) we calculate the energy of the $Q{\bar Q}$ system  
\be
E={1\over 2\pi}\int_{u_0}^{\infty}{b\over {\bar p}} \partial_u x ({u\over
R})^
{7-p} du \ .
\label{E}
\ee
The integral (\ref{E}) is divergent due to the  quark 
self-energy. 
It can be regularized as in
\cite{Malda}:
\bea
E&=&{1\over 2\pi}\; u_0 {1\over 7-p} B({1\over2},{-1\over 7-p})\cr
&&\cr
&=&-{1\over 2\pi}\; u_0 ({1\over 2}-{1\over 7-p})B({1\over2},{6-p\over 7-p}) \ .
\eea
\vskip .5cm

Thus,  
\be
E\sim -\left(\frac{{\bar g}^2_{YM}N\prod_{i~odd}^{2m-1} b_i}{L^2}\right)^{\frac{1}{5-p}} \ .
\label{EE}
\ee
When $p < 5$ the potential is attractive.
When $p=5$ $L$ is independent of $u_0$ and 
the regularized energy is zero. 
In the $p=6$ case
we see that the $Q{\bar Q}$ potential is proportional to $-L^2$ which
results in a repulsive force.
The potential (\ref{EE}) is the same as in the $B=0$ case \cite{BISS} with
$g_{YM}^2 \rightarrow {\bar g}^2_{YM}\prod_{i~odd}^{2m-1} b_i$.
This is presumably  expected by the choice of the moving coordinates frame, and also by the map 
from noncommutative gauge theory to the commutative one \cite{SW}.

\subsection{Non-extremal Dp branes}

In order to compute the expectation value
of the Wilson loop operator in the gauge theory at nonzero temperature
we will use the non-extremal Dp branes background.
We again take the previous string configuration. 
We get
\be
S=\frac{1}{2\pi} \int d\tau du \sqrt{(1-hK^{-1}{\bar p}^2)(1+({u\over R})^{7-p}
hK(\partial_u x)^2)}
+\frac{1}{2\pi}\int d\tau du {{\bar p}\over b}(au)^{7-p}h\partial_u x.
\ee
where $K=1-({u_T\over u})^{7-p}$. 

Solving the equation of motion for $x(u)$ and fixing the end points by 
a constsnt ${{\bar p}\over b}$, we have  
\be
\partial_u x={{\bar p}\over b}({u\over R})^{-{7-p\over2}}K^{-1/2}
\left(({u\over R})^{7-p}K-({{\bar p}\over b})^2\right)^{-1/2} \ .
\ee
Thus,
\be
x(u)=\int_{u_0}^u {{\bar p}\over b}R^{(7-p)}
\left({u}^{7-p}-{u_0}^{7-p}\right)^{-1/2}
\left({u}^{7-p}-{u_T}^{7-p}\right)^{-1/2} \ ,
\ee
wher $u_0$ is the point where the $\partial_u x \rightarrow \infty$,
\be
({au_0})^{7-p}=(au_T)^{7-p}+{\bar p}^2 \ .
\label{u0t}
\ee
Consider two cases:\\

a) Low momentum: $(au_T)^{7-p}\gg {\bar p}^2$. Here the non-extremality effects are
large and we get 
\be
E\sim -\left(\frac{{\bar g}^2_{YM}N\prod_{i~odd}^{2m-1} b_i}
{L^2}\right)^{\frac{1}{5-p}}\left[1+c (T\frac{L^2}{{\bar g}_{YM}^2N\prod_i b_i}
)^{(7-p)/(5-p)}\right] \ ,
\ee 
where $c$ is $N$ independent dimensionless constant.
Again, the potential (\ref{u0t}) is the same as in the $B=0$ case \cite{BISS} with
$g_{YM}^2 \rightarrow {\bar g}^2_{YM}\prod_{i~odd}^{2m-1} b_i$.

b) High momentum: $(au_T)^{7-p}\ll {\bar p}^2$.
Here the noncommutativity effects are large and we get  the noncommutative extremal case
result (\ref{LL}).

\subsection{Wilson Surfaces}

The computation of the expectation value of 
a Wilson surface observable 
amounts in the supergravity approximation to computing  the  minimal 
volume of a membrane bounded at infinity 
by the surface $\Sigma$.
Consider first the wrapped M5 branes background (\ref{m5sol}).

{\it $m=1$}

When the noncommutative effects are large  
the background (\ref{m5sol}) has three small coordinates $x_1,x_2,x_5$.
There are two cases to distinguish. The first is when the membrane wraps one of this
coordinates. In this case the result should coincide with that of
the D4 branes Wilson loop
computation.
The second case is when the membrane is not wrapping one of these small coordinates.
This case is similar to the computation of the potential between monopole and antimonopole.
Here we expect an end fixing problem since unlike
the electric charges in the $B$ field background
there is no useful moving coordinate frame.
 
We start with the first case. 
We denote the membrane coordinates by $\tau,\sigma_1,\sigma_2$.
Consider , for instance, the configuration 
$\tau=x_6, b {\bar R}_0^2\sigma_1=x_5, \sigma_2=u, x_2\equiv x(u)$ and 
$x_1={\bar p}\tau$. 
$\sigma_1$
parametrizes the compactification circle $0\leq \sigma_1 \leq 2\pi$.
The membrane action reads 
\be
S=\frac{1}{(2\pi)^2} \int d\tau d\sigma_1 du \left\{
\sqrt{(1-h^{-1}{\bar p}^2)(1+({u^3\over \pi N b{\bar R}_0^2})
h^{-1}(\partial_u x)^2)}+
{{\bar p}\over b}(au)^{3}h^{-1}\partial_u x\right\} \ ,
\label{M521}
\ee
where here $h = 1+a^3u^3$.
Performing the integration on $\sigma_1$ we
get (\ref{act}) for $p=4$, where $R^3 \rightarrow \pi N b{\bar R}_0^2$. 
This is the expected result.

Consider the second case and let the configuration be
$\tau=x_6,\sigma_1=x_3, \sigma_2=u, x_4\equiv x(u)$. 
Since the $C_{346}$ component is nonzero, the $C$ term in the membrane action
contributes and we get the action per unit length
\be
S=\frac{1}{4\pi^2 b{\bar R}_0^2} \int d\tau du 
\left\{\sqrt{h(1+({u^3\over \pi N b{\bar R}_0^2})(\partial_u x)^2)}+
{a^3\over b}u^3\partial_u x\right\}.
\label{M341}
\ee
The equation of motion for $x(u)$ at large $u$ is of the form
$\partial_u x \sim\; const$, and we have an end fixing problem.
As we noted above, a similar end fixing problem arises when considering the a D2 brane
ending on D4 branes in order to compute the monopole antimonopole potential when $B \neq 0$.

{\it $m=2$}

The computation here is similar to the $m=1$ case when the membrane
is wrapping a small coordinate.
Taking the configuration
$\tau=x_6, b {\bar R}_0^2\sigma_1=x_5, \sigma_2=u, x_2\equiv x(u)$ and 
$x_1={\bar p}\tau$, and integrating the action with respect to
$\sigma$ we get  
(\ref{M521}).

Finally, consider the background (\ref{m5new}).
When the noncommutative effects are large  
the background (\ref{m5new}) has three small coordinates $x_1,x_2,x_5$.
Again we distinguish two types of membrane configuration.
The first is when the  membrane wraps one of this
coordinates. 
A configuration like this is 
$\tau=x_6,\sigma_1=x_1, \sigma_2=u, x_2\equiv x(u)$ and  
$x_1={\bar p}\tau$.
The membrane  action per unit length reads

\be\label{25q}
S=\frac{1}{(2\pi)^2}\int d\tau du \left\{
\sqrt{u^2(1-h^{-1}{\bar p}^2)(1+({u^4\over \pi N})h^{-1}(\partial_u x)^2)}+
{{\bar p}\over b^{3/2}}(au)^{6}h^{-1}\partial_u x\right\},
\ee
where $h=1+a^6u^6$.
The equation of motion for $x(u)$ is 
\be
\frac{u\sqrt{1-h^{-1}{\bar p}^2}h^{-1}{u^4\over \pi N}\partial_u x}
{\sqrt{1+h^{-1}{u^4\over \pi N}(\partial_u x)^2}}+
\frac{{\bar p}}{b^{3/2}}a^6u^6h^{-1}=const \ .
\ee
By choosing the constant to be $\frac{{\bar p}}{b^{3/2}}$, we can fix the end
location of the membrane and we have

\be
\partial_u x=\frac{{\bar p}\;\pi N}{b^{3/2}} u^{-2}(u^6-u_0^6)^{-1/2},
\ee
where $a^6u_0^6={\bar p}^2$.
The distance $L$ which is defined as $x(u\rightarrow\infty)$ reads
\be
L={\sqrt{\pi N}\over u_0}\left({1\over 6}\int_0^1 dy\; (1-y)^{-1/2}y^{-1/3}\right) \ .
\ee
Inserting the solution for $x(u)$ in (\ref{25q}) we get 
the interaction energy per unit length
 between strings of opposite orientation 
\be
E={1\over (2\pi)^2}\int_{u_0}^{\infty}\frac{b^{3/2}}{{\bar p}\pi\;N}\; u^6
\partial_u x\; du \sim
-{N\over L^2} \ .
\ee
This is the same result as for the Wilson surface in the $B=0$
case \cite{Malda}.

The second case is when the membrane is not wrapping one of these small coordinates.
Here we expect an end fixing problem.
Indeed consider the configuration
$\tau=x_6,\sigma_1=x_3, \sigma_2=u, x_4\equiv x(u)$.
The membrane action per unit length reads 

\be
S=\frac{1}{(2\pi)^2} \int d\tau du 
\sqrt{u^2h (1+({u^4\over \pi N})(\partial_u x)^2)}+
{a^6 \over b^{3/2}}u^6\partial_u x \ .
\ee
Writing the equation of motion for $x(u)$ we see that $\partial_u x$
at large $u$ goes like  like $u$ and we have an end fixing problem.

\section{Discussion}

In this  paper we studied the Dp branes supergravity solutions in the presence of a $B$ field,
the decoupling limit and various aspects of the correspondence with the noncommutative
worldvolume field theories.
We analysed the phase structure of the Dp branes and its dependence
on the rank of the $B$ field, i.e.
the dependence  on the number of noncommutative coordinates.
We provided evidence for 
the existence of decoupled Dp branes worldvolume theories  
when $p\geq 6$ 
in presence of a nonzero $B$ field.
The relevance of this to 
M(atrix) theory compactification on the tori $T^p, p \geq 6$ \cite{T6}
in the presence 
of a nonzero $B$ field deserves a further study.
As pointed out \cite{MALR} the D6 branes system has a negative specific heat.
This is usually taken as a sign of instability. 
However, it may be that the noncommutative effects at high energy
require a modification of our field theory understanding 
of thermal equilibrium. This requires further studies too.

We discussed M5 branes in the presence of nonzero $C$ field.
In the case of M5 branes wrapping a circle 
we found the same decoupling limit discussed in \cite{SW}
arising from supergravity. 
In the UV the good description of this system is in terms of D4 branes background,
and we did not find a six dimensional field theory description.
Considering  M5 branes with six flat nocompact worldvolume coordinates we   
found another decoupling limit and we discussed this possible deformation
of the $(0,2)$ SCFT.
We also discussed Type IIB and Type IIA NS5  branes (wrapping a circle)
in the presence of nonzero RR fields.

Finally we computed the expectation value of the Wilson loop (surface) operators
using the dual supergravity description. 
We have seen that, in some cases, in the presence of the nonzero $B$ ($C$) field 
there is a way to fix the string (membrane) end point (string) 
by considering a moving coordinates frame in the computation.
The results for both extremal and non-extremal 
Dp branes (and for the M5 branes) are the same as in the $B=0$ case with
$g_{YM}^2 \rightarrow {\bar g}^2_{YM}\prod_{i~odd}^{2m-1} b_i$.
This is presumably  expected by the map 
from noncommutative gauge theories to the commutative ones \cite{SW}.

\section*{Acknowledgements}

We would like to thank J. Maldacena, S. Yankielowicz and D. Youm for 
useful discussions. M.A. is supported by the John Bell Scholarship from
the World Laboratory.

\end{document}